# The String Model for Ultrarelativistic Nuclear Scattering [*]

**K. WERNER** [†]

Institut für Theoretische Physik, Universität Heidelberg
Philosophenweg 19, 69120 Heidelberg, Germany [‡]

**Abstract**

We review basic theoretical concepts to describe nuclear collisions at ultrarelativistic energies. We discuss relativistic strings, Gribov–Regge theory (GRT) of hadronic interactions and string models based on GRT, and finally generalizations to nucleus–nucleus scattering.

## 1   Introduction

The early universe was probably a hot and dense "fireball" of quarks and gluons, before, due to expansion and cooling, hadrons emerged. Presently, there are considerable efforts to create such a "quark gluon plasma" (QGP) in nucleus–nucleus scattering at ultrarelativistic energies ($\gg 1$ GeV per nucleon) [1].

In ultrarelativistiv collisions there is certainly enough energy available to produce high enough energy densities for a QGP, provided all the energy is used to heat up the system. This is, however, not the case. We know that the nuclei are to some extent transparent, they go through each other by keeping a large fraction of their original momentum. But still, the nucleons do loose energy, which shows up as baryon–poor matter in the central region. To be more precise: the system has roughly the form of

---

a cylinder which expands essentially longitudinally, the forward and backward front moving almost with the velocity of light. The forward and backward region of the cylinder are baryon–rich, the central region is baryon–poor. On the other hand, the energy density is largest in the central region, and the big question is whether the energy density in this region is large enough to form a plasma.

Theoretically, such questions related to the space–time structure can be investigated by using string models [2]–[6]. These models provide a full description of ultrarelativistic nucleus–nucleus collisions, starting really with two incident nuclei and not at some vaguely known intermediate stage as many hydrodynamical models. Basic features of high energy hadronic collisions as tranparency, longitudinal structure ... can be easily understood in terms of string models, even more, these are basic properties of string models and not just a specific choice of parameters. String models are also very useful to study signals of the QGP, since practically all signals are strongly affected by the space–time evolution of the system.

There is quite a number of microscopic dynamical models, a fact which is at first sight somewhat disturbing. Looking closer, one finds, however, similarities. There is a whole class of models (VENUS [2], the dual parton model (DPM) [3, 4], the quark gluon string model (QGSM) [5, 6]) with all the models being strictly based on Gribov-Regge theory and Veneziano's cylinder hypothesis (these models are referred to as Gribov–Regge models, GRM's). This general framework is referred to as "the string model" in this article. The different models of this class (VENUS, DPM, QGSM) differ in details concerning the precise formulation of string formation and decay. A nice feature of the Gribov–Regge approach is the consistent formulation within relativistic quantum theory. The classical string is only used as a phenomenological parametrization of particle production, this is not a classical model.

There is no real alternative to the string model discussed in this article. There is a "classical string model" [7], which seems to be quite different but nevertheless successful. However, also here the hadronic interaction amounts to string formation and decay. Lacking theoretical guidance one simly assumes something about string formation, which is actually very similar to the "colour exchange mechanism" of the GRM's. Another successful approach is the RQMD model [8], which uses classical trajectories and measured hadron–hadron cross sections. But also here, for energetic hadron–hadron interactions, a string approach is applied. Furthermore, at high energies, the Gribov cross sections are used to introduce multiple scattering. So all successful models use at least elements of GRT, even when they are not formulated within this framework.

In this article we treat only the class of models based on Gribov–Regge theory (VENUS, DPM, QGSM), the general framework referred to as "the string model". We discuss the theoretical concepts as relativistic strings and Gribov–Regge theory, and demonstrate how both are linked to provide "the string model of hadronic interactions". We also discuss the generalization to nucleus–nucleus scattering.Many details and references missing in this article can be found in a recent review article

[2].

## 2 String Dynamics

The dynamics and fragmentation of relativistic strings are crucial ingredients of the model for hadronic interactions to be introduced later. We shall discuss strings in a general fashion in order to demonstrate that dynamics and fragmentation are almost fixed from symmetry requirements. This is the reason that string fragmentation models are much less arbitrary than one might think.

We review classical string theory [9, 10, 11]. We discuss how to obtain a string action, we derive equations of motion for the space-time evolution of strings, and we discuss the general solution as well as conservation laws. We then treat the simplest possible string solution, the so called "yo-yo" string.

### 2.1 A Gauge Invariant String Action

We discuss in this subsection how to obtain a string action from invariance requirements.

A classical string is a two-dimensional surface in the four-dimensional Minkowski space,

$$x = x(\tau, \sigma) ,\tag{1}$$

with a spacelike parameter $\sigma$ and a timelike one $\tau$. Of course this is only one of infinitely many parametrizations of this surface. A transformation

$$\begin{pmatrix}\tau \\ \sigma\end{pmatrix} \longrightarrow \begin{pmatrix}\tilde{\tau}(\tau,\sigma) \\ \tilde{\sigma}(\tau,\sigma)\end{pmatrix} \tag{2}$$

from one parameter space to another is called a *gauge transformation*, and the group of such transformations is called a *gauge group*. One assumes that the string action should not depend on the parametrization, so gauge invariance is a necessary requirement. Further restrictions should be locality and covariance. Concerning the question of gauge invariance, it is useful to relate a metric $g$ to a certain string parametrization via

$$g_{\alpha\beta} = \partial_\alpha x^\mu \partial_\beta x_\mu , \tag{3}$$

where $\alpha$ and $\beta$ assume the values 1 and 2, and where we used $\partial_1 \equiv \frac{\partial}{\partial \tau}$ and $\partial_2 \equiv \frac{\partial}{\partial \sigma}$. By taking "dot" and "prime" as abbreviations for $\frac{\partial}{\partial \tau}$ and $\frac{\partial}{\partial \sigma}$, the metric can be written as

$$g = \begin{pmatrix} \dot{x}\dot{x} & \dot{x}x' \\ x'\dot{x} & x'x' \end{pmatrix} . \tag{4}$$

How does the metric $g$ transform under the gauge transformations given in eq. (2)? Defining the two component variable $\xi$ via

$$\xi_1 \equiv \tau , \quad \xi_2 \equiv \sigma , \tag{5}$$

and using
$$\tilde{\partial}_\alpha \equiv \frac{\partial}{\partial \tilde{\xi}_\alpha} \,, \tag{6}$$

we get
$$\begin{align} g_{\alpha\beta} &= \partial_\alpha x^\mu \partial_\beta x_\mu \tag{7} \\ &= \tilde{\partial}_i x^\mu \partial_\alpha \tilde{\xi}_i \tilde{\partial}_j x_\mu \partial_\beta \tilde{\xi}_j \tag{8} \\ &= \partial_\alpha \tilde{\xi}_i \tilde{g}_{ij} \partial_\beta \tilde{\xi}_j \,. \tag{9} \end{align}$$

Since the components of the Jacobi matrix $M$ of the gauge transformation eq. (2) are given as
$$M_{ab} = \partial_b \tilde{\xi}_a \,, \tag{10}$$

we can write eq. (9) in matrix notation as
$$g = M^T \tilde{g} M \,. \tag{11}$$

This leads to the identity
$$\sqrt{|\det g|} = \sqrt{|\det \tilde{g}|} \, |\det M| \,. \tag{12}$$

On the other hand we have
$$d^2 \tilde{\xi} = |\det M| \, d^2 \xi \,, \tag{13}$$

which together with eq. (12) immediately suggests that a $\xi$ integration over $\sqrt{|\det g|}$ is invariant under gauge transformations,
$$\tilde{I} \equiv \int \sqrt{|\det \tilde{g}|} \, d^2 \tilde{\xi} = \int \sqrt{|\det g|} \, d^2 \xi \equiv I \,. \tag{14}$$

Writing the integral $I$ explicitly as
$$I = \int \sqrt{(x'\dot{x})^2 - x'^2 \dot{x}^2} \, d\tau d\sigma \tag{15}$$

shows that $I$ is also local and covariant. In fact, $I$ is the simplest local, covariant, and gauge invariant expression, and so $I$ is a very attractive candidate for a string action. Therefore we define the action of a relativistic string to be [9]
$$S = \int L \, d\tau d\sigma \,, \tag{16}$$

with
$$L = -\kappa \sqrt{-\det g} = -\kappa \sqrt{(x'\dot{x})^2 - x'^2 \dot{x}^2}, \tag{17}$$

where we used $|\det g| = -\det g$. We will see later that the proportionality constant $\kappa$ can be identified with the "string tension", the energy per unit length of the string.

## 2.2 Equations of Motion, Conservation Laws

We rewrite the action defined in the last subsection more explicitly as

$$S = \int_{\tau_1}^{\tau_2} d\tau \int_0^\pi d\sigma \, L \, , \tag{18}$$

with

$$L = -\kappa \sqrt{(x'\dot{x})^2 - x'^2 \dot{x}^2} \, . \tag{19}$$

We use the convention $\sigma_{\min} = 0$ and $\sigma_{\max} = \pi$, where $\pi$ is an arbitrary number at the moment. The symbols $\tau_1$ and $\tau_2$ represent initial and final times. To obtain the equations of motion, we require

$$\delta S = 0 \tag{20}$$

under infinitesimal variations $\delta x(\sigma, \tau)$ of the string surface. We find the equations of motion

$$\frac{\partial}{\partial \tau} \frac{\partial L}{\partial \dot{x}_\mu} + \frac{\partial}{\partial \sigma} \frac{\partial L}{\partial x'_\mu} = 0 \, , \tag{21}$$

and the boundary conditions

$$\frac{\partial L}{\partial x'_\mu} = 0 \quad \text{at } \sigma = 0, \pi \, . \tag{22}$$

From the invariance of the action under translations one obtains conservation laws for energy and momentum. Defining the energy–momentum currents as

$$P_\tau^\mu := -\frac{\partial L}{\partial \dot{x}_\mu} \, , \qquad P_\sigma^\mu := -\frac{\partial L}{\partial x'_\mu} \, , \tag{23}$$

we may introduce the string momentum in various ways, for example as

$$P^\mu(\text{string}) := \int_{\mathcal{C}_\tau} d\sigma \, P_\tau^\mu \, , \tag{24}$$

where integration at constant $\tau$ is implied. From $\delta S = 0$, one concludes that the string momentum is conserved.

Using the currents defined in eq. (23), we may rewrite the equations of motion eq. (21) as

$$\frac{\partial}{\partial \tau} P_\tau^\mu + \frac{\partial}{\partial \sigma} P_\sigma^\mu = 0 \, , \tag{25}$$

and the boundary condition eq. (22) reads

$$P_\sigma^\mu = 0 \, , \quad \text{at } \sigma = 0, \pi \, . \tag{26}$$

Our next aim will be to solve these equations of motion.

## 2.3  Solutions of the String Equations

To solve the equations of motion, we choose a gauge which simplifies the equations of motion. The orthonormal gauge

$$x'\dot{x} = 0 , \qquad \dot{x}^2 + x'^2 = 0 \tag{27}$$

does so. The currents eq. (23) are now

$$P_\tau = \kappa \dot{x} , \qquad P_\sigma = -\kappa x' , \tag{28}$$

the equations of motion eq. (25) are simply wave equations

$$\ddot{x}_\mu - x''_\mu = 0 , \tag{29}$$

and the boundary conditions eq. (26) are

$$x'(t,0) = x'(t,\pi) = 0 . \tag{30}$$

To completely specify the gauge, we set

$$x_0 \equiv t = \tau . \tag{31}$$

This implies that $\tau$ and $\sigma$ have length dimensions. For the following we use this *lab frame parametrization*, and consider only the space components of $x$. Eq. (27) now reads

$$\vec{x}'\dot{\vec{x}} = 0, \qquad (\dot{\vec{x}})^2 + (\vec{x}')^2 = 1 . \tag{32}$$

The solution of eqs. (29, 30) is

$$\vec{x}(t,\sigma) = \frac{1}{2}[\vec{y}(t+\sigma) + \vec{y}(t-\sigma)] , \tag{33}$$

where $\vec{y}(t)$ is obviously the trajectory of one endpoint, $\vec{x}(t,0)$, called the directrix. The directrix has to be periodic,

$$\vec{y}(t + 2\pi) - \vec{y}(t) = \frac{2\vec{P}}{\kappa} , \tag{34}$$

where $\vec{P}$ is the string momentum (to be shown later). Eq. (33) has the following meaning: each point on the string may be obtained by a simple geometrical construction once the directrix $\vec{y}(t)$ is known.

## 2.4  The Yo-Yo String

A simple but important example is the so-called yo-yo string, characterized by a one-dimensional directrix with one period consisting of two linear segments. In case of a one-dimensional directrix, straight lines with a tilt of 45° against vertical (in space-time) are mandatory, because the string end (represented by $\vec{y}(t)$) moves with the velocity of light (because of eqs. (32, 30)). From eq. (33) it is clear that the corresponding string is a simple straight line ($q\bar{q}$ in fig. 2) stretched between directrix and antidirectrix.

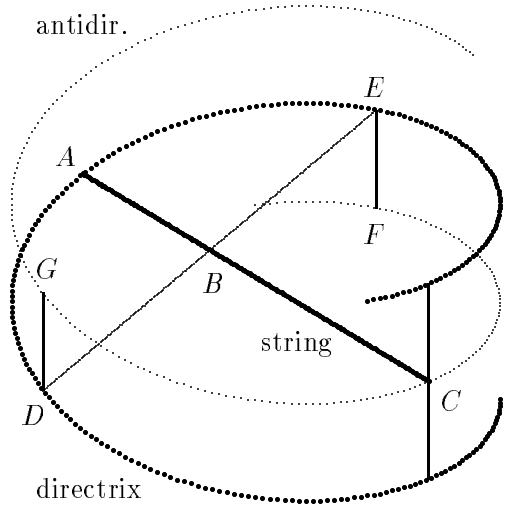

**Figure 1**: A string with its directrix and antidirectrix. The directrix segment $DAE$ defines the string piece $AB$ and the antidirectrix $FCG$ the string piece $CB$.

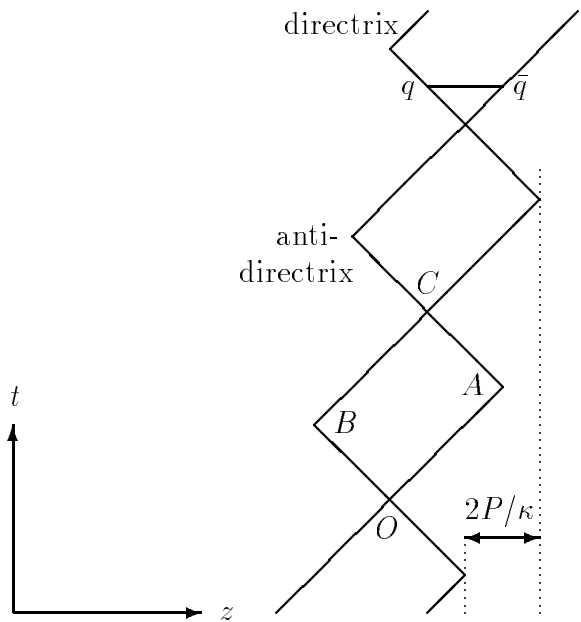

**Figure 2**: Space-time picture of a "yo-yo" string with period $2P/\kappa$.

## 3　String Fragmentation

We discuss the rules for string breaking in the framework of classical relativistic string theory in general and in particular for yo-yo strings. Although, in classical string theory, the time evolution is fixed once a string breakpoint is known, the determination of locations of breakpoints requires further input. For this purpose we employ the same symmetry arguments which led earlier to the string action. This procedure leads to the "area law" [12, 13, 14], i.e. the probability $dP$ of a string to

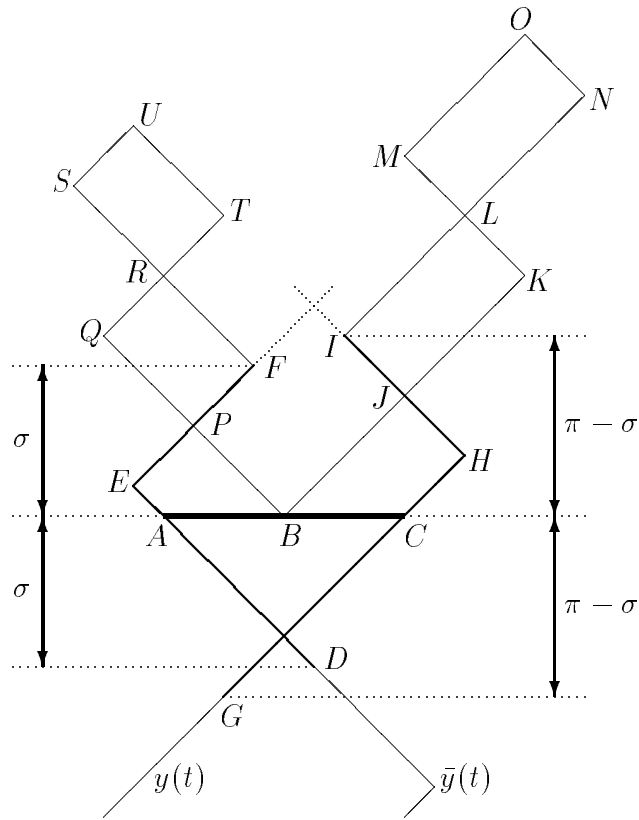

**Figure 3**: Breaking of a "yo-yo" string: we first determine the segments of the (anti)directrix corresponding to the string pieces $AB$ and $CB$; these segments are then periodically continued into the future.

break within a given area element $d^2A$ is given as

$$dP = (1-P)\,\alpha\,d^2A\,,\tag{35}$$

with the "break probability" $\alpha$ as a parameter. Knowing the breakpoint, it is clear know how to proceed. As for the action, we assume locality. If a break occurs at $x(t,\sigma)$, we have to make sure that for the future as well as the past we have periodic (anti-) directrices, and that the directrices for future and past match properly in the present. The only way to ensure this is to periodically continue the directrix corresponding to one string piece and the antidirectrix corresponding to the other string piece into the future. This fully determines the time evolution of either string piece also for all the future, at least untill the next break.

Let us now discuss these "cutting rules" for a yo-yo string (see fig. 3). Without interaction, the string stretches between directrix $(t, y(t))$ and antidirectrix $(t, \bar{y}(t)) = (t, x(t, \pi))$. Let the point $B = (t, x(t, \sigma))$ be a breakpoint on the string at time $t$, dividing the string into two segments $AB$ and $BC$ with $A = (t, x(t, 0))$ and $C = (t, x(t, \pi))$. The directrix and antidirectrix corresponding to these segments are

$DEF$ with

$$D = (t - \sigma, y(t - \sigma)),  \tag{36}$$
$$F = (t + \sigma, y(t + \sigma))  \tag{37}$$

and $GHI$ with

$$G = (t - (\pi - \sigma), \bar{y}(t - (\pi - \sigma))),  \tag{38}$$
$$I = (t + (\pi - \sigma), \bar{y}(t + (\pi - \sigma))).  \tag{39}$$

Using

$$\bar{y}(t) = y(t - \pi) + \frac{P}{\kappa},  \tag{40}$$

we verify easily that, after the appropriate shift, the segments $DEF$ and $GHI$ provide a full period of the unperturbed string. As discussed earlier, we obtain the directrices of the two segments after the break by continuation of $DEF$ ($\to DEFS\cdots$) and of $GHI$ ($\to GHIN\cdots$). The corresponding antidirectrices can be easily constructed from the relation

$$\bar{y}(t) = \frac{1}{2}(y(t + \pi) + y(t - \pi))  \tag{41}$$

between directrix $y$ and antidirectrix $\bar{y}$, and we get $BKM\cdots$ and $BQT\cdots$. We realize the identities

$$\|BJ\| = \|JK\|, \quad \|HJ\| = \|JI\|  \tag{42}$$

and

$$\|BP\| = \|PQ\|, \quad \|EP\| = \|PF\|,  \tag{43}$$

which provide a very simple procedure for actually constructing the new directrices in numerical applications.

## 4  Gribov–Regge Theory (GRT)

Gribov's multiple scattering theory of ultrarelativistic hadronic interactions, referred to as Gribov–Regge theory (GRT), is the theoretical basis of the string model of hadronic/nuclear scattering to be introduced later.

We briefly discuss the Pomeron. We then introduce an expression for the amplitude of elastic hadron–hadron scattering due to multiple Pomeron exchange. We calculate the elastic and, via the optical theorem, the total cross section. We discuss the AGK cutting rules, which provide a technique to calculate the inaginary part of elastic amplitudes. Finally we apply the AGK rules to expand the total cross section as $\sigma_{\mathrm{tot}} = \sum \sigma_m$, with "topological cross sections" $\sigma_m$ referring to $m$ elementary inelastic processes ($m$ cut Pomerons).

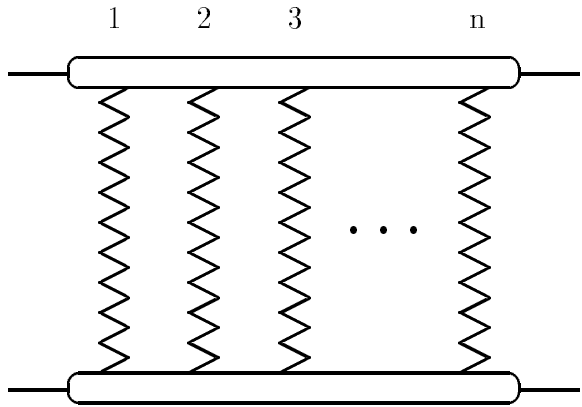

**Figure 4**:
$n$ Pomeron exchange.

## 4.1 The Pomeron

In the next subsection we introduce a multiple scattering theory of hadron–hadron interactions, implying multiple exchange of elementary objects called Pomerons. From general considerations of the high energy limit of elastic amplitudes, one parametrizes the amplitude associated with Pomeron exchange as

$$A(s,t) \sim s^{\alpha(t)} \approx s^{\alpha(0)+\alpha't}, \qquad (44)$$

with $s$ and $t$ being the Mandelstam variables $s = (p_1 + p_2)^2$ and $t = (p_1 - p_3)^2$, with $p_1$, $p_2$ ($p_3$, $p_4$) being the momenta of the incoming (outgoing) hadrons.

The nature of the Pomeron in terms of quarks and gluons is still not known. Initially, Pomerons were thought to be ladder diagrams (gluon ladders), or gluon networks of cylindrical topology, but QCD calculations are not conclusive. We adopt Venezianos picture of a Pomeron being a cylinder of gluons and quark loops (being a generalized gluon ladder).

## 4.2 The Multi–Pomeron Amplitude

Starting point is the following expression for the elastic amplitude:

$$A_{2\to 2}(s,t) = \sum_n A_n(s,t) , \qquad (45)$$

with

$$A_n(s,t) = \frac{i^{n-1}\pi^{1-n}}{n!} \int \prod_{i=1}^n d^2k_i \, \delta^{(2)}(k - \sum k_i) \, N_n(k_1 \cdots k_n) \, D(s, k_1^2) \cdots D(s, k_n^2) , \qquad (46)$$

representing $n$ Pomerons exchanges (fig. 4) . The variables $k$ and $k_i$ represent transverse momenta. The Pomeron Green's function is

$$D(s, k^2) = \eta \left(\frac{s}{s_0}\right)^{\alpha(-k^2)-1} . \qquad (47)$$

Introducing a *rapidity gap* $y$ via $y = \ln s/s_0$ and defining

$$\Delta := \alpha(0) - 1 , \tag{48}$$

we find

$$D(s, k^2) \approx \eta \, \exp(\Delta y) \, \exp\left[-\alpha'(0)\, y\, k^2\right] , \tag{49}$$

with $\exp(\Delta y)$ being 1, if we have an ideal Pomeron with $\alpha(0) = 1$. As we are going to argue later, the data, however, suggest that $\alpha(0)$ is slightly larger than 1. The simple form of the Pomeron Green's function is a pure assumption (of Regge pole dominance), only justified by the success of the theory, and it is, therefore, most important to justify this assumption within QCD.

Eqs. (46, 49) are the basis for applications to be discussed in the following.

## 4.3 Elastic Scattering and Total Cross Section

As a first application of the Gribov–Regge theory, we consider elastic scattering. The amplitude for elastic scattering is given as

$$A(s, t) = \sum_{n=1}^{\infty} A_n(s, t) , \tag{50}$$

where $A_n$ represents $n$ Pomeron exchanges, and is given by eqs. (46, 49). Assuming factorization of the vertex function,

$$N_n(k_1, \cdots, k_n) = C^{n-1} \prod_{i=1}^{\infty} N(k_i^2) , \tag{51}$$

we get

$$A(s, t) = \frac{i}{4\pi} \int d^2b \, \exp(i\vec{k}\vec{b}) \, \gamma(s, b) , \tag{52}$$

with

$$\gamma(s, b) = \frac{1}{C} \left\{ 1 - \exp\left[-C\,\omega(s, b)\right] \right\} , \tag{53}$$

and

$$\omega(s, b) = \frac{N_0 \, \exp(\Delta y)}{R^2 + \alpha' y} \exp\left[-\frac{b^2/4}{R^2 + \alpha' y}\right] . \tag{54}$$

Eqs. (52, 53, 54) may be used to calculate cross sections as

$$\sigma_{\text{tot}} = 8\pi \, \text{Im} A(s, 0) = \int d^2b \, 2 \, \text{Re}\gamma \tag{55}$$

and

$$\sigma_{\text{el}} = \int dk^2 \, 4\pi \, |A|^2 = \int d^2b \, |\gamma|^2 . \tag{56}$$

There are five free parameters, $N_0$, $\Delta$, $R^2$, $\alpha'$, and $C$, which are fixed by comparing with data. The increase of $\sigma_{\text{tot}}(s)$ and $\sigma_{\text{in}}(s)$ with $s$ as well as many elastic scattering data can be nicely reproduced, not so for $\Delta = 0$.

## 4.4 The Abramovskiĭ–Gribov–Kancheli Cutting Rules

Due to the optical theorem, the discontinuity (or imaginary part) of the elastic amplitude is related to inelastic processes. One therefore investigates disc$A$ in order to study inelastic scattering. The Abramovskiĭ–Gribov–Kancheli (AGK) Cutting Rules [15] provide a technique to express disc$A$ (or Im$A$) in terms of disc$G$, representing elementary inelastic processes associated with the exchange of a single Pomeron.

We start with the elastic amplitude for multiple Pomeron exchange, given in eq. (46), which we write as

$$i\, A_n = \int d\Omega \prod_{\gamma=1}^{n} i\, G_\gamma \;, \tag{57}$$

with $G_i \equiv G(s, k_i^2) = N(k_i^2) D(s, k_i^2)$. The Cutkoski cutting rules state for a Feynman diagram (or a sum of graphs) having the structure eq. (57):

$$\frac{1}{i}\mathrm{disc}A_n = \sum_{\mathrm{cuts}} \int d\Omega \prod_{\substack{\mathrm{left}\\ \mathrm{of\ cut}}} i\, G_\alpha \prod_{\substack{\mathrm{right}\\ \mathrm{of\ cut}}} \frac{1}{i} G_\beta^* \prod_{\substack{\mathrm{cut}\\ \mathrm{Pomerons}}} \frac{1}{i}\mathrm{disc}G_\mu = \sum_m A_{nm}\;, \tag{58}$$

where $m$ represents the number of elementary inelastic interactions ("cut Pomerons"). One finds analytic expressions for $A_{nm}$ [2].

## 4.5 An Expansion of the Total Cross Section

We are now going to use the expansion of $\frac{1}{i}\mathrm{disc}A_n$ in terms of the number of cut Pomerons to obtain an expansion of the total cross section in terms of *topological cross sections*: $\sigma_{\mathrm{tot}} = \sum \sigma_m$. Here $\sigma_m$ corresponds to the cross section of $m$ elementary inelastic processes ($m$ cut Pomerons). We have

$$\sigma_{\mathrm{tot}}(s) = \frac{1}{2is}\mathrm{disc}T(s,0) = \frac{4\pi}{i}\mathrm{disc}A(s,0) \tag{59}$$

$$= 4\pi \sum_{n=1}^{\infty} \frac{1}{i}\mathrm{disc}A_n(s,0)\;, \tag{60}$$

where $n$ reflects $n$ Pomeron exchanges. We obtain

$$\sigma_{\mathrm{tot}}(s) = \sum_{m=0}^{\infty} \sigma_m(s)\;, \tag{61}$$

with

$$\sigma_m(s) = \frac{8\,\pi\, N_0\, \exp(\Delta y)}{m\, z}\Big[1 - e^{-z}\sum_{k=0}^{m-1}\frac{z^k}{k!}\Big] \quad ; m > 0\;, \tag{62}$$

which may be evaluated easily numerically. For $m = 0$, we get

$$\sigma_0(s) = 8\,\pi\, N_0\, \exp(\Delta y) \sum_{n=1}^{\infty} \frac{1 - 2^{n-1}}{n\, n!}\left(-\frac{z}{2}\right)^{n-1}\;. \tag{63}$$

The $\sigma_m$ are strictly positive!

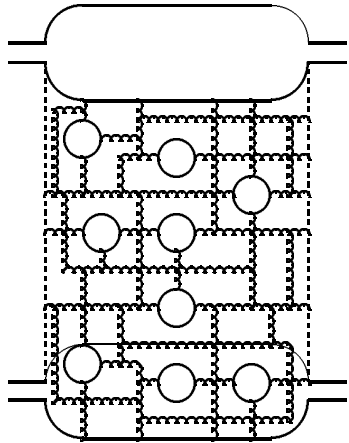 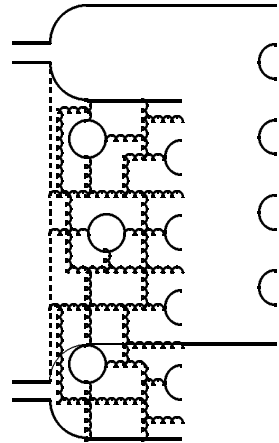

**Figure 5**: Cylinder diagram (gluons and quark loops on the back sheet are not drawn).

**Figure 6**: Cut cylinder diagram (gluons and quark loops on the back sheet are not drawn).

## 5 The VENUS String Model

The Gribov–Regge theory provides a framework to calculate amplitudes $A_{2\to 2}$ for elastic hadron–hadron scattering, and therefore elastic and total cross sections can be calculated. For more detailed investigations, one needs inelastic amplitudes $A_{2\to n}$, describing particle production. These amplitudes are, however, not calculable within GRT. On the other hand, we know that discontinuities (or imaginary parts) of elastic amplitudes are related to inelastic scattering, so one may take the expansion of $\mathrm{disc} A_{2\to 2}$ (or of $\sigma_{\mathrm{tot}}$) from the last section as a guideline to construct a model. This is exactly what is done in the VENUS model:

- In VENUS, elastic amplitudes are calculated strictly according to Gribov–Regge Theory (GRT), and so are the elastic and total cross sections.

- VENUS provides a model for calculating inelastic amplitudes, guided by the expansion of $\sigma_{\mathrm{tot}}$ in terms of topological cross sections.

The VENUS model to be introduced in the following is closely related to the dual parton model (DPM) [3, 4], introduced by Capella et al., and the quark gluon string model (QGSM) by Kaidalov et al. [5, 6].

In order to formulate a model for inelastic scattering, one needs to know something about the nature of the Pomeron. According to Veneziano [16], a Pomeron is a cylinder, see fig. 5. Correspondingly we identify the discontinuity of the Pomeron propagator (or more precisely $-i\,\mathrm{disc}G = 2\,\mathrm{Im}G$) with a squared cut cylinder, with a cut cylinder shown in fig. 6. The two cutting edges of the cut cylinder are identified with relativistic strings, as shown in fig. 7, where the thin lines represent (anti)quarks and the thick lines socalled remnants, each one representing an incident hadron minus the corresponding (anti)quark. So, before the interaction, (anti)quark and remnant of the same hadron are connected, after the interaction, (anti)quark and remnant from different hadrons. Therefore the term "colour exchange" is used, because it looks like colour being exchanged between the two (anti)quarks.

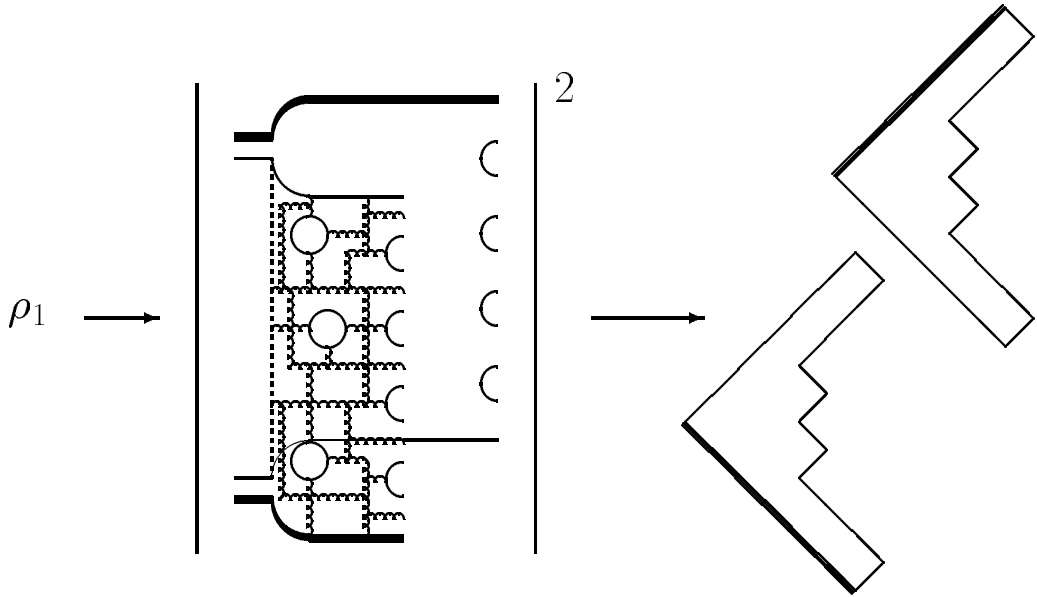

**Figure 7**: Cut cylinder corresponding to two strings.

Considering two incoming hadrons $f_1(\tilde{p}_1)$, $f_2(\tilde{p}_2)$, with $f(p)$ refering to flavour (momentum), and a colour exchange between the (anti)quarks $i(k)$ and $j(l)$, one obtains two strings $S_+$ and $S_-$. The string

$$S_+ := \left\{ \tilde{f}_1(\tilde{p}_1) - i(k) \, \middle| \, j(l) \right\} \tag{64}$$

contains at one end the projectile remnant $\tilde{f}_1(\tilde{p}_1) - i(k)$, being the projectile $\tilde{f}_1$ reduced by the parton (quark or antiquark) $i$. This string end carries the projectile momentum $\tilde{p}_1$ reduced by the parton momentum $k$. The other end consists of the parton $j$ with momentum $l$. Since this string contains the projectile remnant, which has a large forward momentum, we refer to this string as forward string or forward baryonic string (it carries baryon number 1). The other string

$$S_- := \left\{ \tilde{f}_2(\tilde{p}_2) - j(l) \, \middle| \, i(k) \right\} \tag{65}$$

contains at one end the target remnant $\tilde{f}_2(\tilde{p}_2) - j(l)$, which is the target $\tilde{f}_2$ reduced by the parton $j$, with the target momentum $\tilde{p}_2$ reduced by the parton momentum $l$. On the other end, we have the parton $i$ with momentum $k$. This string is referred to as backward string or backward baryonic string (it carries as well baryon number 1). So if, for example, $\tilde{f}_1$ and $\tilde{f}_2$ are protons and $i$ and $j$ are $u$ quarks, the two strings are $ud$–$u$ strings, with the diquark $ud$ being in one case the forward end and in the other case the backward end.

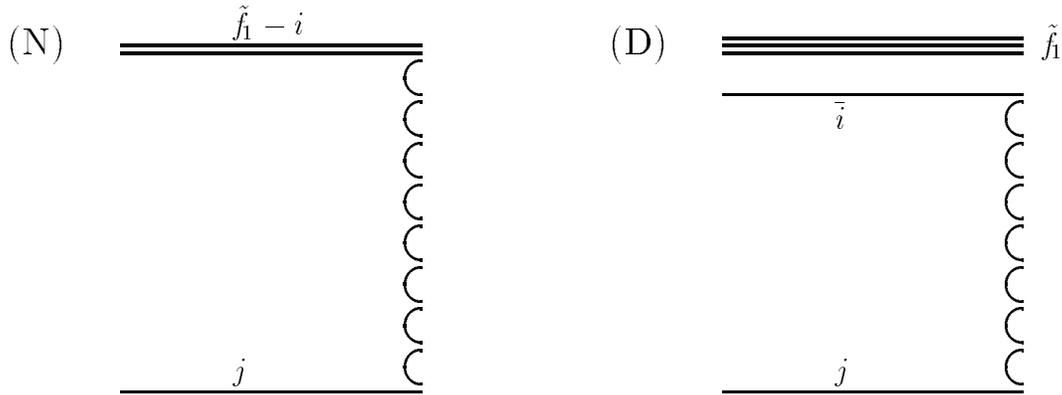

**Figure 8**: Nondiffractive (N) and diffractive (D) type.

The forward string $S_+$ consists of a backward parton $j$ and a forward hadron $\tilde{f}$ minus a parton $i$. It might be by chance that $\tilde{f}_1 - i$ is equal to $\tilde{f}_1 + \bar{i}$ with $\tilde{f}_1$ having survived as a singlet, or, in other words, the CE involved only a colour singlet $i - \bar{i}$ pair and left $\tilde{f}_1$ surviving as a spectator. The same arguments apply of course for the backward string $S_-$. We refer to such interactions as "diffractive" (D) and to others as "nondiffractive" (N) (see fig. 8). We say, the hadron suffers a D–type or N–type interaction (we use weights $w$ and $1 - w$).

For the full 1–cylinder contribution (consisting of two strings), we have four combinations, NN, ND, DN, and DD, with weights $(1-w)^2$, $(1-w)w$, $w(1-w)$, and $w^2$, which are referred to as

$$\begin{array}{ll} \text{nondiffractive scattering} & \text{(NN)}, \\ \text{diffractive projectile excitation} & \text{(ND)}, \\ \text{diffractive target excitation} & \text{(DN)}, \\ \text{Pomeron–Pomeron scattering} & \text{(DD)}. \end{array} \qquad (66)$$

These contributions are conveniently represented as quark–line diagrams as shown in fig. 9.

## 6  Nucleus-Nucleus Interactions

We are going to generalize the results of the preceeding sections to nucleus-nucleus ($A$–$B$) scattering in a straightforward way.

For elastic scattering, we consider diagrams of the type shown in fig. 10, where the dashed line with index $n$ represents a block of $n$ Pomeron exchanges. The same nucleon may (or may not) be involved in more than one interaction. We write the

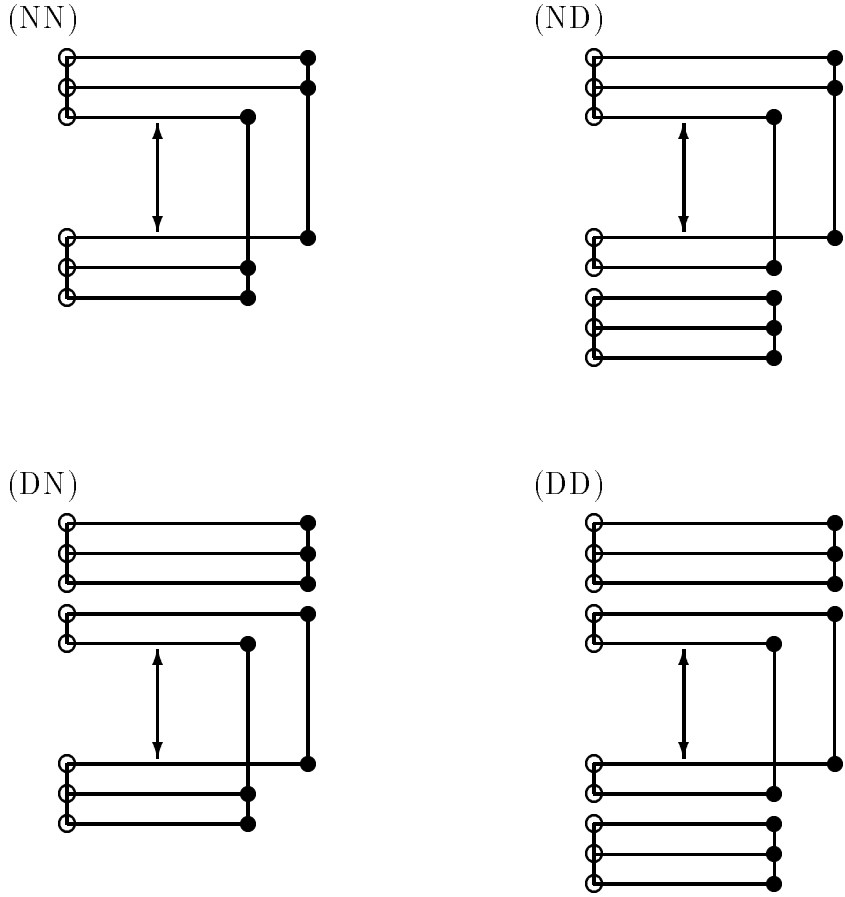

**Figure 9**: Quark line diagrams representing the four contributions to single colour exchange between quarks.

elastic amplitude for $A$–$B$ scattering as

$$A = \sum_{\nu} \sum_{\alpha_1 \beta_1 \ldots \alpha_\nu \beta_\nu} \sum_{n_1 \ldots n_\nu} A^{\alpha_1 \beta_1 \ldots \alpha_\nu \beta_\nu}_{n_1 \ldots n_\nu} , \qquad (67)$$

with

$$i A^{\alpha_1 \beta_1 \ldots \alpha_\nu \beta_\nu}_{n_1 \ldots n_\nu} = \int d\Omega \, N^{\alpha_1 \beta_1 \ldots \alpha_\nu \beta_\nu}_{n_1 \ldots n_\nu} \prod_{\mu=1}^{\nu} \frac{1}{n_\mu!} \prod_{j=1}^{n_\mu} i \, D , \qquad (68)$$

where $\sum_{\alpha_1 \beta_1 \ldots}$ represents a sum over all possible "collision sequences" $\alpha_1 \beta_1 \ldots \alpha_\nu \beta_\nu$, with $\alpha_\mu$ and $\beta_\mu$ being indices refering to the projectile and target nucleon involved in the $\mu^{th}$ interaction. Ordering in the sequence is irrelevant, and the double indices have to be pairwise different: $\alpha_i \beta_i \neq \alpha_j \beta_j$. Eq. (68) corresponds to a sequence of $\nu$ N–N collisions, each of them representing a "block" of several pomeron exchanges (see fig. 10). Now we expand the absorptive part of $i A^{\alpha_1 \beta_1 \ldots \alpha_\nu \beta_\nu}_{n_1 \ldots n_\nu}$ as a sum of terms

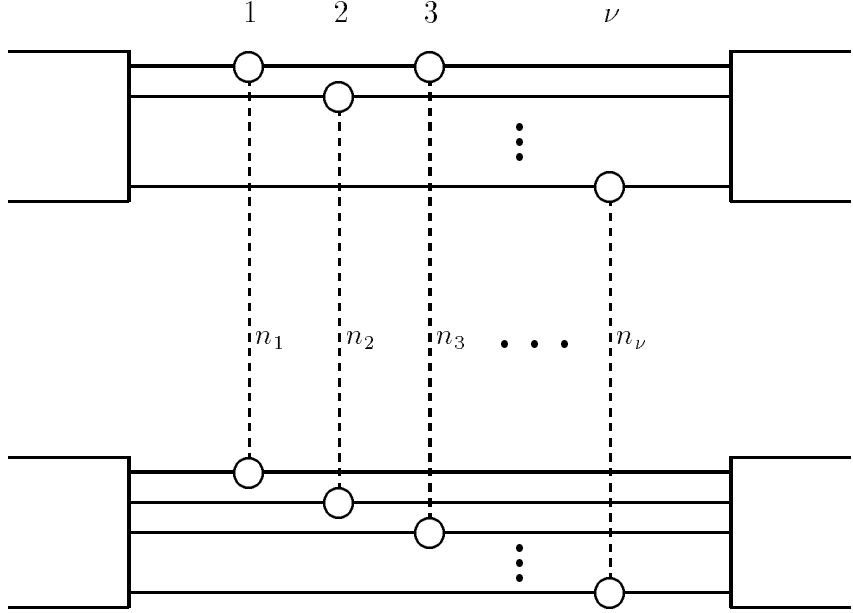

**Figure 10**: Elastic nucleus-nucleus scattering: nucleons of the projectile nucleus interacts with nucleons from the target nucleus via a "block" of several Pomeron exchanges (dashed line) for each $N$–$N$ interaction.

corresponding to $m_\mu$ out of the $n_\mu$ Pomerons being cut,

$$2\,\mathrm{Im} A^{\alpha_1\beta_1...\alpha_\nu\beta_\nu}_{n_1...n_\nu} = \frac{1}{i}\mathrm{disc} A^{\alpha_1\beta_1...\alpha_\nu\beta_\nu}_{n_1...n_\nu} = \sum_{m_1...m_\nu} A^{\alpha_1\beta_1...\alpha_\nu\beta_\nu}_{n_1...n_\nu,m_1...m_\nu} \qquad (69)$$

(see fig. 11), with

$$A^{\alpha_1\beta_1...\alpha_\nu\beta_\nu}_{n_1...n_\nu,m_1...m_\nu} = \sum_{I_{\mathrm{el}} I_{\mathrm{ib}}} \int d\Omega\, N^{\alpha_1\beta_1...\alpha_\nu\beta_\nu}_{n_1...n_\nu}$$

$$\prod_{\mu \in I_{\mathrm{in}}} (-1)^{n_\mu - m_\mu} \binom{n_\mu}{m_\mu} \frac{1}{n_\mu!} \prod_{j=1}^{n_\mu} 2\,\mathrm{Im} D$$

$$\prod_{\mu \in I_{\mathrm{el}}} (-1)^{n_\mu} \frac{1}{n_\mu!} \left[ \prod_{j=1}^{n_\mu} 2\,\mathrm{Im} D - 2\,\mathrm{Re} \prod_{j=1}^{n_\mu} \frac{1}{i} D \right]$$

$$\prod_{\mu \in I_{\mathrm{ib}}} (-1)^{n_\mu} \frac{1}{n_\mu!} 2\,\mathrm{Re} \prod_{j=1}^{n_\mu} \frac{1}{i} D\ . \qquad (70)$$

$I_{\mathrm{in}}$, $I_{\mathrm{el}}$, and $I_{\mathrm{ib}}$ are sets of indices, corresponding to inelastic, elastic, and inter–block cuts: $I_{\mathrm{in}}$ is the set of indices $\mu$ with $m_\mu > 0$, and $I_{\mathrm{el}}$, $I_{\mathrm{ib}}$ are partitions of $I_\nu \setminus I_{\mathrm{in}}$, with $I_\nu = \{1...\nu\}$, which means $I_{\mathrm{el}} \cup I_{\mathrm{ib}} = I_\nu \setminus I_{\mathrm{in}}$. The sum $\sum_{I_{\mathrm{el}} I_{\mathrm{ib}}}$ is meant to sum over all partitions. Using the optical theorem, one obtains

$$\sigma^{AB}_{\mathrm{in}} = \int d^2 b\, \tilde{\sigma}^{AB}_{\mathrm{in}}(b)\ , \qquad (71)$$

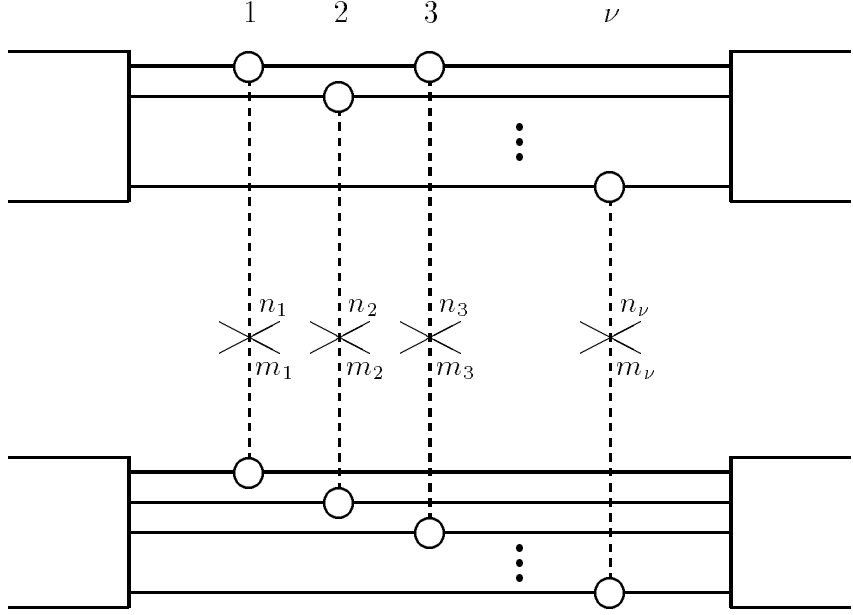

**Figure 11**: A contribution to the absorptive part of the elastic amplitude for nucleus-nucleus scattering, related to cutting $m_\mu$ out of the "block" of $n_\mu$ exchanged Pomerons in the $\mu^{th}$ nucleon-nucleon interaction.

with

$$\tilde{\sigma}_{\text{in}}^{AB}(b) = \sum_{\nu>0} \sum_{\alpha_1\beta_1...\alpha_\nu\beta_\nu} \sum_{m_1...m_\nu} \tilde{\sigma}_{m_1...m_\nu}^{\alpha_1\beta_1...\alpha_\nu\beta_\nu}(b) \, , \tag{72}$$

$$\tilde{\sigma}_{m_1...m_\nu}^{\alpha_1\beta_1...\alpha_\nu\beta_\nu}(b) = \int dT_{AB} \prod_{\mu=1}^{\nu} \xi_{m_\mu}(b_\mu) \prod_{\mu=\nu+1}^{AB} [1 - \xi(b_\mu)] \, , \tag{73}$$

$$dT_{AB} := \prod_{\alpha=1}^{A} d^2 b_\alpha^A \, T(b_\alpha^A) \prod_{\beta=1}^{B} d^2 b_\beta^B \, T(b_\beta^B) \, , \tag{74}$$

$$b_\mu := b - b_{\alpha_\mu}^A + b_{\beta_\mu}^B \, , \tag{75}$$

$$\xi(b) := \frac{1}{C} \left\{ 1 - \exp\left[-2C\omega(b)\right] \right\} \, , \tag{76}$$

$$\xi_m(b) := \frac{1}{C} \frac{\left[2C\omega(b)\right]^m}{m!} \exp\left[-2C\omega(b)\right] \, . \tag{77}$$

Performing the sum in eq. (72), one obtains

$$\tilde{\sigma}_{\text{in}}^{AB}(b) = 1 - \int dT_{AB} \prod_{\mu=1}^{AB} [1 - \xi(b_\mu)] \, . \tag{78}$$

This formula, derived in the GRT framework without approximations (apart from assuming a simple form of the nucleus–nucleon vertex) is the starting point of many applications. Eq. (78) is often referred to as the Glauber formula, being a generalization of Glauber's result for $h$–$A$ scattering, which has been derived, however, within nonrelativistic scattering theory. Eq. (78) is still very complicated, being a $AB$–fold integration, which cannot be reduced to a product of integrations as in the $h$–$A$ case. Eqs. (73) and (78) are our basic formulas for the Monte Carlo treatment, which allow to generate appropriate collision sequences and thus to simulate nucleus–nucleus scattering.